\documentclass[12pt]{article}
\usepackage{latexsym}
\usepackage{amsmath,,calrsfs}
\usepackage{amsfonts}
\usepackage{amssymb}
\usepackage{amscd}
\usepackage{bbm}
\usepackage{fancybox}
\usepackage{cite}
\usepackage{amsmath,amsfonts,amsbsy}
\usepackage{pstricks,pst-node}
\usepackage[small,bf,hang]{caption2}
\usepackage{graphicx}
\usepackage{epsfig}
\usepackage{psfrag}
\usepackage{comment}
\usepackage{youngtab}
\usepackage{young}

\usepackage{float}

\psset{unit=1.3cm,linewidth=.5pt,radius=.2}  

\usepackage{multirow}                     
\usepackage{float}                          
\usepackage{lscape}                         
\usepackage{bm}

\newcommand{\beq}{\begin{equation}}
\newcommand{\eeq}{\end{equation}}
\newcommand{\bi}{\begin{itemize}}
\newcommand{\ei}{\end{itemize}}
\newcommand{\bt}{\begin{tabular}}
\newcommand{\et}{\end{tabular}}
\newcommand{\bc}{\begin{center}}
\newcommand{\ec}{\end{center}}

\newcommand{\be}{\begin{equation}}
\newcommand{\ee}{\end{equation}}
\newcommand{\bea}{\begin{eqnarray}}
\newcommand{\eea}{\end{eqnarray}}
\newcommand{\ba}{\begin{array}}
\newcommand{\ea}{\end{array}}

\def\bbox{{\,\lower0.9pt\vbox{\hrule \hbox{\vrule height 0.2 cm
\hskip 0.2 cm \vrule height 0.2 cm}\hrule}\,}}
\newcommand{\dsl}{\pa \kern-0.5em /}

\makeatletter \@addtoreset{equation}{section} \makeatother

\def\slashchar#1{\setbox0=\hbox{$#1$}           
   \dimen0=\wd0                                 
   \setbox1=\hbox{/} \dimen1=\wd1               
   \ifdim\dimen0>\dimen1                        
      \rlap{\hbox to \dimen0{\hfil/\hfil}}      
      #1                                        
   \else                                        
      \rlap{\hbox to \dimen1{\hfil$#1$\hfil}}   
      /                                         
   \fi}

\begin{document}

\begin{titlepage}
\begin{center}

\hfill UG-12-23

\vskip 1.5cm

{\Large \bf  On Topologically Massive  Spin-2 Gauge Theories\\[.3truecm] beyond Three Dimensions
}

\vskip 1cm

{\bf Eric A.~Bergshoeff\,, Marija Kovacevic\,,
Jan Rosseel and Yihao Yin }\\

\vskip 25pt

{\em Centre for Theoretical Physics,
University of Groningen, \\ Nijenborgh 4, 9747 AG Groningen, The
Netherlands \vskip 5pt }

{email: {\tt E.A.Bergshoeff@rug.nl, maikovacevic@gmail.com,
j.rosseel@rug.nl, y.yin@rug.nl}} \\

\end{center}

\vskip 0.5cm

\begin{center} {\bf ABSTRACT}\\[3ex]
\end{center}

We investigate in which sense, at the linearized level, one can extend the
3D topologically massive gravity theory beyond three dimensions. We show that, for each $k=1,2,3,\cdots$ a
free topologically massive  gauge theory in $4k-1$ dimensions can be defined describing a massive ``spin-2'' particle  provided one uses a non-standard representation of the massive ``spin-2'' state which makes use of a
two-column Young tableau where each column is of height $2k-1$. We work out the case of $k=2$, i.e.~7D, and show, by canonical analysis,
that the model describes, unitarily, 35 massive ``spin-2'' degrees of freedom. The issue of
interactions is discussed and compared with the three-dimensional situation.

\end{titlepage}

\newpage
\setcounter{page}{1} \tableofcontents

\newpage

\section{Introduction}

Three-dimensional higher-derivative theories of gravity have received considerable attention over the years. The first example of such a higher-derivative theory  is  the ``Topologically Massive Gravity''  (TMG) model \cite{Deser:1981wh}.
The TMG Lagrangian consists  of the usual Einstein-Hilbert (EH) term, which by itself does not describe any degrees of freedom in three dimensions, and a Lorentz Chern-Simons (LCS) term which is parity-odd and third-order in the derivatives. The two terms together describe a single  massive state of
helicity +2 or --2, depending on the relative sign between the EH and LCS terms. A more recent example is the   ``New Massive Gravity'' (NMG) model\cite{Bergshoeff:2009hq}.
NMG is the parity even version of TMG and its Lagrangian contains besides the EH term a particular  combination of two fourth-order derivative terms, of which one is quadratic in the Ricci tensor and the other   is quadratic in the Ricci scalar. The NMG Lagrangian describes, unitarily,  two massive
  states of helicity +2 and --2. The signs in front of the kinetic terms corresponding to these two states  are the same as a
consequence of the fact that the Lagrangian is parity even.

Recently, it was pointed out that the NMG model can be extended to four dimensions,  at the linearized level, provided one describes the massive spin-2 state by a non-standard representation
 corresponding to a mixed-symmetry Young tableau
with two columns of height 2 and 1, respectively \cite{Bergshoeff:2012ud}.
A similar extension does not apply to the TMG model. This can be understood as follows.
One may view TMG as the ``square root'' of NMG in the same way that one may view Topologically Massive Electrodynamics (TME) \cite{Siegel:1979fr}
as the ``square root'' of the Proca theory.
The latter property is based on the fact that the Klein-Gordon operator, when acting on divergence-free vectors, as it does in the 3D Proca equation, factorises into the product of two first-order operators each of which separately describes a single  state of helicity $+1$ and $-1$  \cite{Bergshoeff:2009tb}.\,\footnote{Alternatively, one may act on vectors that are {\sl not} divergence-free. The product of the two first-order operators then leads to a modified Proca equation. Next, by taking the divergence of this modified equation one may derive that the vector is divergence-free.}
The equation of motion describing one of the two helicity states is a massive self-duality equation
\cite{Townsend:1983xs,Deser:1984kw}.
This property of the 3D Proca equation carries over to the 3D Fierz-Pauli (FP) equation, describing masssive
spin-2 particles, where the Klein-Gordon operator acts on
a divergence-free symmetric tensor of rank 2. It also applies to 3D generalised FP equations, describing massive
 particles of higher spin, where one considers
symmetric tensors of rank $p>2$ \cite{Bergshoeff:2011pm}.

The above property of the Klein-Gordon operator, when acting on 3D divergence-free vectors,
can be extended as follows. Consider a generalized Proca equation where the Klein-Gordon
operator acting on a divergence-free form-field of  given rank gives zero. One can show that
in $D=4k-1$ dimensions this Klein-Gordon operator
factorizes into the product of two first-order operators provided the form-field is of rank $2k-1$.
Each of the two operators describes half of the helicity states that were described by the original generalized
Proca equation. For $k=1$ one obtains 3D 1-forms which we already discussed. The next case to consider is $k=2$
which leads to 3-forms in D=7 dimensions. The corresponding massive self-duality equation was encountered first
in the  context of seven-dimensional gauged supergravity where the mass $m$
plays the role of the gauge coupling constant \cite{Townsend:1983xs}. The 7D Proca equation describes
20 degrees of freedom that transform as the ${\bf 10}^+ + {\bf 10}^-$ of the little group SO(6).
The ${\bf 10}^+$ and ${\bf 10}^-$ degrees of freedom are each separately described by the two massive self-duality
equations.\,\footnote{A similar factorisation of the Klein-Gordon operator, when acting on
divergence-free 5D 2-forms, requires that one
considers a Klein-Gordon operator with the wrong sign in front of the mass term \cite{Townsend:1983xs}.
Such a wrong sign can be avoided by considering a symplectic doublet of 2-forms and using the corresponding epsilon
symbol in the massive self-duality equation. This is very similar to extending Majorana spinors to
Symplectic Majorana spinors. We will not consider this possibility
further in this letter.}

As we will discuss in this letter the above property of the 7D Proca equation carries over to
generalised FP equations \cite{Curtright:1980yk,Labastida:1987kw,Bekaert:2002dt} in $D=4k-1$
dimensions where the Klein-Gordon operator acts on fields whose
indices are described by a GL(D,$\mathbb{R}$) Young tableau with an arbitrary number of columns each of which has
height $2k-1$. We are interested in models describing propagating massive
spin-2 particles that generalize, at the linearized level, the 3D
TMG model.\,\footnote{A different extension, which we will not consider here, is to add higher-derivative topological terms to the
Einstein-Hilbert term. Such an extension in 7D has been considered in \cite{Lu:2010sj}.}
 Interpreting  ``spin''  in higher dimensions as the
number of columns in the Young tableau that characterizes the index
structure of the field under consideration,\,\footnote{More precisely, for massless spins we
 only consider two-column Young tableaux where the first column has a maximum number
of D$-$3 boxes. For massive spins the maximum number is D$-$2. The Young tableaux with more boxes describe either ``spin 1'' particles
or no degrees of freedom at all.} we are led to consider
7D fields $h_{\mu_1\mu_2\mu_3,\nu_1\nu_2\nu_3}$ whose index
structure is given by the following GL(7,$\mathbb{R})$ Young
tableau
\begin{equation} \label{youngsymm}
 \begin{tabular}{l}{\footnotesize
\begin{Young}
$\mu_{1}$ & $\nu_{1}$ \cr
 $\mu_{2}$ & $\nu_{2}$ \cr
 $\mu_{3}$ & $\nu_{3}$ \cr
\end{Young}
}
\end{tabular} \,.
\end{equation}
In order to keep in line as much as possible with the construction
of the 3D TMG model, and, furthermore, to avoid writing down too
many indices, we will use a notation where  $\bar{\mu}$ stands for a
collection of three antisymmetrized indices $\mu_1$, $\mu_2$ and
$\mu_3$, i.e.~$\bar{\mu} \ \leftrightarrow \ [\mu_1\,  \mu_2 \,
\mu_3]$ or $h_{\bar\mu,\bar\nu} \equiv
h_{\mu_1\mu_2\mu_3,\nu_1\nu_2\nu_3}$. If we regard the field $h$ as
a field describing the propagation of a massive particle via a
generalised FP equation, the number of propagating degrees of
freedom equals the dimension of the irreducible, traceless, representation of the little
group SO(6), given by the  same Young diagram \eqref{youngsymm}.
This leads to 70 propagating degrees of freedom which transform as
the $\mathbf{35}^+ + \mathbf{35}^-$ of SO(6). These two
representations are interchanged by the action of parity.

In the next section we wish to construct a parity violating free 7D ``Topologically Massive Spin-2 Gauge Theory'' for the field $h$, such that 35 degrees of freedom are propagated. This theory is an analogue of
the 3D TMG model at the linearized level. The construction of this topologically massive gauge theory  will proceed in the same fashion as
can be done for the 3D  TMG model. We will first consider the massive self-duality equation and, next, boost up the number of derivatives by solving the differential subsidiary conditions.

\section{The Model}

Our starting point are the generalised FP equations for a field ${\tilde h}$ with the symmetry properties
\eqref{youngsymm}. These equations consist of the Klein-Gordon equation
\begin{equation} \label{FP}
(\Box - m^2) {\tilde h}_{\bar{\mu},\bar{\nu}} = 0 \,,
\end{equation}
together with two subsidiary constraints, one algebraic and one differential:
\begin{equation}\label{FP2}
\eta^{\mu \nu} {\tilde h}_{\bar{\mu},\bar{\nu}}  =  0 \,,\hskip 3truecm
\partial^\mu {\tilde  h}_{\bar{\mu},\bar{\nu}}  =  0 \,.
\end{equation}
We have used here a notation where the contraction of an unbarred index $\mu$ with a barred index $\bar{\mu}$ means that the index $\mu$ is contracted with the first index $\mu_1$ of the collection $\bar{\mu}$, e.g.
\begin{equation}
\partial^\mu {\tilde  h}_{\bar{\mu},\bar{\nu}} = \partial^{\mu_1} {\tilde  h}_{\mu_1\mu_2 \mu_3,\nu_1\nu_2\nu_3} \,.
\end{equation}
Note that the symmetry properties of ${\tilde  h}$ imply that divergence-freeness on the first three indices of ${\tilde h}$ also implies divergence-freeness on the second three indices. One can show via an explicit counting that the two subsidiary constraints  reduce the number of components of ${\tilde h}$ to 70 propagating degrees of freedom.

To obtain a massive self-duality equation for ${\tilde  h}$ we use the property that  the Klein-Gordon operator
$(\Box - m^2) \delta_{\bar{\mu}}^{\bar{\nu}}$ in the space of divergence-free 1-forms can be factorized as follows
\begin{equation}
(\Box - m^2) \delta_{\bar{\mu}}^{\bar{\nu}} = \left(
\frac{1}{3!}\varepsilon _{\bar{\mu} }{}^{\alpha \bar{\rho}}\partial
_{\alpha }+m\delta _{\bar{\mu}}^{\bar{\rho} }\right) \left(
\frac{1}{3!}\varepsilon _{\bar{\rho}}{}^{\beta \bar{\nu} }\partial
_{\beta }-m\delta _{\bar{\rho}}^{\bar{\nu}}\right)\,.
\end{equation}
This suggests the following massive self-duality equation for
${\tilde h}$:
\begin{equation}  \label{sqrtFP}
\left( \frac{1}{3!}\varepsilon _{\bar{\mu}}{}^{\alpha
\bar{\rho}}\partial _{\alpha }-m\delta
_{\bar{\mu}}^{\bar{\rho}}\right) {\tilde  h}_{\bar{\rho},\bar{\nu} }=0  \,.
\end{equation}
A similar massive self-duality equation describing the parity transformed
degrees of freedom is obtained by replacing $m$ by $-m$.
Contracting the massive self-duality equation \eqref{sqrtFP} with $\partial^\mu$
leads to the divergence-freeness condition of ${\tilde h}$.
Furthermore, a contraction with $\eta^{\mu \nu}$ of the same
equation and using the symmetry properties of ${\tilde h}$ proofs
the tracelessness condition of ${\tilde h}$. The Schouten identity
shows that the tensor
$\varepsilon_{\bar{\mu}}{}^{\alpha\bar{\rho}}\partial_{\alpha }
{\tilde h}_{\bar{\rho},\bar{\nu}}$ has the same symmetry properties
as ${\tilde  h}$ provided that  ${\tilde  h}$ is divergence-free and
traceless.

We next proceed by  boosting up the derivatives of the above model
by solving the differential subsidiary condition that expresses that
${\tilde h}$ is divergence-free, see eq.~\eqref{FP2}.
This condition is solved in terms of a new field $h$, with the same index structure and symmetry properties as $h$,
by applying twice the Poincar\'e lemma for 3-forms:
one time on the $\bar\mu$ indices  of ${\tilde h}_{\bar\mu,\bar\nu}$ and a second time on the $\bar\nu$ indices  of ${\tilde h}_{\bar\mu,\bar\nu}$.
One thus obtains the following solution
\begin{equation}\label{solsub}
{\tilde h}_{\bar\mu,\bar\nu} = G_{\bar\mu,\bar\nu}(h)\,,
\end{equation}
where the tensor $G_{\bar\mu,\bar\nu}(h)$ is  defined by
\begin{equation} \label{diffinv}
 G_{\bar{\mu},\bar{\nu}}( h) =\varepsilon _{\bar{\mu}}{}^{\alpha
\bar{\rho}}\varepsilon _{\bar{\nu}}{}^{\beta \bar{\sigma}}\partial _{\alpha
}\partial _{\beta }\,h_{\bar{\rho},\bar{\sigma}}\,.
\end{equation}
Using a Schouten identity, one can  show that the tensor $G(h)$ has the same symmetry properties as $h$.
In terms of $h$ the massive self-duality
equation now reads
\begin{equation}\label{hdsdm}
\left( \frac{1}{3!}\varepsilon _{\bar{\mu}}{}^{\alpha
\bar{\rho}}\partial _{\alpha }-m\delta
_{\bar{\mu}}^{\bar{\rho}}\right) G_{\bar{\rho},\bar{\nu} }(h)=0\,.
\end{equation}

We note that the higher-derivative equations of motion in terms of $h$ are  invariant under gauge transformations of $h$
 with a gauge parameter $\xi$ that has a
symmetry structure corresponding to a Young tableau with two
columns, one of height 3 and one of height 2. Schematically, in
terms of Young tableaux, these gauge transformations are
given by, ignoring indices, $\delta h = \partial \xi$ or, in terms of Young tableaux, by
\begin{equation}\label{schem}
\delta\hskip -.3truecm
\begin{tabular}{l}{\footnotesize
 \begin{Young}
 & \cr
 & \cr
 & \cr
\end{Young}
}
\end{tabular}
\  =\
\begin{tabular}{l}{\footnotesize
 \begin{Young}
 & \cr
 & \cr
 & $\partial$\cr
\end{Young}
}
\end{tabular}\,.
\end{equation}
It is understood here that when taking the derivative of the gauge parameter at the right-hand-side  one first takes  the curl of
the two indices in the second column of the Young tableau describing the index structure  of the gauge parameter,
and next applies a Young symmetrizer\,\footnote{A Young symmetrizer is an operator that projects onto the symmetries corresponding to
a given Young tableaux.  For the precise definition  and its basic properties,
see e.g.~\cite{Fulton,Hamermesh}. Following the notation of
\cite{Francia:2005bv} a Young symmetrizer $Y_{[p,q]}$ is a
projection operator, $Y^2=Y$, that acts on a $(p,q)$ bi-form and
projects onto the part that corresponds to a two-column Young
tableau of height $p$ and $q$, respectively. When the bi-form is
already of the desired symmetry type it acts like the identity
operator. For instance, $Y_{[3,3]} h_{\bar\mu,\bar\nu} =
h_{\bar\mu,\bar\nu}$. }
to obtain the same
index structure at both sides of the equation.

The gauge-invariant
curvature $R(h)$ of $h$ is obtained by hitting $h$
with two derivatives: one which takes the curl of the first three
indices of $h$ and another which takes the curl of the second
three indices:
\begin{equation}
R_{\alpha\bar\rho,\beta\bar\sigma}(h) = \partial_{[\alpha}\partial^{[\beta} h_{\bar\rho],}{}^{\bar\sigma]}\,.
 \end{equation}
 This leads to a curvature tensor with an index
structure corresponding to a Young tableau with two columns of
height 4. By construction, this curvature tensor satisfies a
generalised Bianchi identity. The
tensor $G(h)$ defined above is obtained from the curvature $R(h)$  by taking the dual on the first 4 indices of $R(h)$
and a second dual on the second 4 indices. One thus obtains a tensor  corresponding to a
Young tableau with two columns of height 3 each.
Due to the Bianchi identity of $R(h)$, the  tensor $G(h)$ is divergence-free in each of its indices. We therefore call it the
``Einstein tensor'' of $h$.

Summarizing we have

\begin{equation}
{ h}\   = \hskip -.3truecm \begin{tabular}{l}{\footnotesize
\begin{Young}
  &  \cr
  &  \cr
  &  \cr
\end{Young}
}
\end{tabular}
\ \ \  \rightarrow  \ \ \
R(h)\   = \hskip -.3truecm  \begin{tabular}{l}{\footnotesize
\begin{Young}
  &  \cr
  &  \cr
  &  \cr
  $\partial$&$\partial$\cr
\end{Young}
}
\end{tabular}
\ \ \  \rightarrow  \ \ \ G(h)\ =\ {}^\star {}^\star R(h)\ = \hskip -.3truecm \begin{tabular}{l}{\footnotesize
\begin{Young}
  &  \cr
  &  \cr
  &  \cr
\end{Young}
}
\end{tabular}\,.
\end{equation}

The equations of motion \eqref{hdsdm} for $h$ describe the same degrees of freedom as the original massive self-duality equation
\eqref{sqrtFP}  for ${\tilde h}$. For instance,
the trivial solution ${\tilde h}=0$ of the massive self-duality equation \eqref{sqrtFP}  is mapped under eq.~\eqref{solsub} to
the solutions of the equation $G_{\bar\mu,\bar\nu}(h)=0$\,.
Since the Einstein tensor  $G(h)$ is the double dual of the curvature
$R(h)$  this equation implies that the curvature of $h$ is zero.
This in its turn implies that $h$ is a pure gauge degree of freedom \cite{Bekaert:2002dt}.

The equations of motion \eqref{hdsdm} define  a 7D Topologically Massive Spin-2 Gauge Theory.
We note that these equations imply that the Einstein tensor of $h$ is traceless, i.e.~$\eta^{\mu\nu}G_{\bar\mu ,\bar\nu}(h)=0$.
To construct an action giving rise to these equations it is useful to introduce the following ``generalized Cotton tensor'':
\begin{equation} \label{Ctensor}
C_{\bar{\mu},\bar{\nu}}(h) = Y_{[3,3]} \left[ \varepsilon
_{\bar{\mu}}{}^{\alpha \bar{\rho}}\partial _{\alpha }G_{
\bar{\rho},\bar{\nu}}\left( h\right) \right]  \,,
\end{equation}
where $Y_{[3,3]}$ is a Young symmetrizer,
that ensures that $C_{\bar{\mu},\bar{\nu}}$
has the  symmetry properties of the Young tableau given in
eq.~\eqref{youngsymm}. Note that we have to write this Young
symmetrizer explicitly, as we want to use the Cotton tensor in the
action and we cannot assume that the  condition that
$G_{\bar{\mu},\bar{\nu}}$ is  traceless is satisfied off-shell. Once
one can show that, as a consequence of the equations of motion, $G$
is traceless, the Young symmetrizer can be dropped. Independent of
whether $G_{\bar\mu,\bar\nu}$ is traceless or not, one can show that
the Cotton tensor $C_{\bar{\mu},\bar{\nu}}$ is  divergence-free on
both sets of indices $\bar{\mu}$ and $\bar{\nu}$, as well as
traceless
\begin{equation}\label{constraint}
\partial^\mu C_{\bar{\mu},\bar{\nu}} = 0 \,, \qquad \eta^{\mu \nu} C_{\bar{\mu},\bar{\nu}} = 0 \,.
\end{equation}

The equations of motion \eqref{hdsdm} can now be integrated to the
following action:\,\footnote{Note that, due to the second constraint
in \eqref{constraint}, the first term in \eqref{7Daction} has a
generalized scale invariance. This is similar to the scale
invariance of the 3D Cotton tensor.}
\begin{equation}\label{7Daction}
I\left[h\right] =\int d^{7}x\left\{
\frac{1}{12}h^{\bar{\mu},\bar{\nu}}C_{ \bar{\mu},\bar{\nu}}\left(
h\right) -\frac{1}{2}mh^{\bar{\mu},\bar{\nu}}G_{
\bar{\mu},\bar{\nu}}\left( h\right) \right\} \text{ .}
\end{equation}
This action defines the 7D Topologically
Massive  Spin-2 Gauge Theory.
Indeed, varying this action with respect to $h$ leads to the equations of motion
\begin{equation}
\frac{1}{6}C_{\bar{\mu},\bar{\nu}}\left( h\right)
-m G_{\bar{\mu},\bar{\nu} }\left( h\right) =0\text{ .}  \label{TMG
original eom}
\end{equation}
Contracting these equations of motion with $\eta^{\mu \nu}$, one  obtains the tracelessness condition
\begin{equation}
\eta^{\mu\nu}G_{\bar{\mu},\bar{\nu}}\left( h\right) =0\text{ .}
\end{equation}
With the tracelessness condition in hand,  the Young
symmetrizer in (\ref{Ctensor}) can be dropped, and  the
equation of motion (\ref {TMG original eom}) reproduces the equation of motion given in
eq.~\eqref{hdsdm}.

\section{Canonical Analysis}

As a check we will verify, by canonical analysis, that the action \eqref{7Daction} indeed describes 35 spin-2 degrees of freedom.
We first split the indices into temporal and spatial components like $\mu = (0,i)\,, i=1,\cdots,6,$ and impose the gauge-fixing conditions
\begin{equation}\label{gcondition}
\partial ^{i}h_{i\mu _{2}\mu _{3},\nu _{1}\nu _{2}\nu _{3}}=0\text{ .}
\end{equation}
We next parametrize  $h$ in terms of the independent components $(a,b,c,d,e)$ as follows:\,\footnote{
The notation $\left\{ \ \right\} _{\rm a.s.}$ stands for antisymmetrizing all indices within  the curly bracket that have  the same latin letter. For instance, $\left\{ S_{i_2 i_3 j_1 j_2 j_3} \right\} _{\rm a.s.}=  S_{[ i_2 i_3 ] [ j_1 j_2 j_3 ]}  $.
}
\begin{subequations}
\label{canonical decomp.}
\begin{eqnarray}
h_{0i_{2}i_{3},0j_{2}j_{3}} &=&a_{i_{2}i_{3},j_{2}j_{3}}\text{ ,} \\ [.2truecm]
h_{0i_{2}i_{3},j_{1}j_{2}j_{3}} &=&\varepsilon
_{j_{1}j_{2}j_{3}}{}^{k_{1}k_{2}k_{3}}\partial
_{k_{1}}b_{k_{2}k_{3},i_{2}i_{3}}\ + \ 
\left\{ \left( \delta _{i_{3}j_{3}}-\frac{\partial _{i_{3}}\partial _{j_{3}}
}{\nabla ^{2}}\right) c_{j_{1}j_{2},i_{2}}\right.  \notag \\ [.2truecm]
&&\left. \ \ \ \ +\left( \delta _{i_{2}j_{2}}\delta
_{i_{3}j_{3}}-\frac{\partial _{i_{2}}\partial _{j_{2}}}{\nabla
^{2}}\delta _{i_{3}j_{3}}-\delta _{i_{2}j_{2}}\frac{\partial
_{i_{3}}\partial _{j_{3}}}{\nabla ^{2}}\right)
d_{j_{1}}\right\}_{\rm a.s.} \text{ ,} \\ [.2truecm]
h_{i_{1}i_{2}i_{3},j_{1}j_{2}j_{3}} &=&\varepsilon
_{i_{1}i_{2}i_{3}}{}^{k_{1}k_{2}k_{3}}\varepsilon
_{j_{1}j_{2}j_{3}}{}^{l_{1}l_{2}l_{3}}\partial _{k_{1}}\partial
_{l_{1}}e_{k_{2}k_{3},l_{2}l_{3}}\text{ .}
\end{eqnarray}
\end{subequations}
All components $a,b,c,d,e$ are divergence-free. Furthermore, the components $b,c$  are traceless in each pair of its indices but the components $a$
and $e$  contain their traces.

It is instructive to count the different degrees of freedom at this
point. Our starting point is the field $h$ of symmetry-type
\eqref{youngsymm} which is in the ${\bf 490}$ representation of
GL(7,$\mathbb{R}$). This field transforms under the gauge
transformations schematically denoted by \eqref{schem}. We should be
careful with counting the number of independent gauge parameters
because the gauge transformations \eqref{schem} are double
reducible: the 490 gauge parameters $\xi$ have their own gauge
symmetry with 210 gauge parameters $\zeta$ which are given by, ignoring indices,  $\delta\xi  = \partial\zeta$ or in terms
of Young tableaux by
\begin{equation}\label{schemg2}
\delta\hskip -.3truecm
\begin{tabular}{l}{\footnotesize
 \begin{Young}
 & \cr
 & \cr
  \cr
\end{Young}
}
\end{tabular}
\  =\
\begin{tabular}{l}{\footnotesize
 \begin{Young}
 & \cr
 & $\partial$\cr
  \cr
\end{Young}
}
\end{tabular}\,.
\end{equation}
In its turn the 210 gauge parameters $\zeta$ have a gauge symmetry
by themselves with 35 gauge parameters $\lambda$ which are
irreducible. These transformations are given by, ignoring indices,  $\delta\zeta=\partial\lambda$ or in terms of Young tableaux by
\begin{equation}\label{schemg3}
\delta\hskip -.3truecm
\begin{tabular}{l}{\footnotesize
 \begin{Young}
 & \cr
  \cr
  \cr
\end{Young}
}
\end{tabular}
\  =\
\begin{tabular}{l}{\footnotesize
 \begin{Young}
 & $\partial$\cr
  \cr
  \cr
\end{Young}
}
\end{tabular}\,.
\end{equation}
A correct counting yields that there are 490--210+35 = 315
independent gauge parameters. The gauge symmetries corresponding to these gauge parameters are fixed by
the gauge conditions \eqref{gcondition} on the field $h$. To see this, one first varies \eqref{gcondition}
under the $\xi$-symmetries \eqref{schem} and requires this variation to be zero. The resulting condition
on the $\xi$-parameters has a gauge-symmetry which can be fixed by imposing the following restriction on the $\xi$-parameters:
\begin{equation}
\partial ^{i_{2}}\xi _{i_{2}\mu _{3},\nu _{1}\nu _{2}\nu _{3}} =0\text{ .}
\label{gauge-fix zeta0}
\end{equation}
Varying this condition under the $\zeta$-symmetries \eqref{schemg2} leads to a gauge-invariant condition on the $\zeta$-parameters.
To fix this gauge symmetry we impose
the following gauging-fixing conditions on the $\zeta$-parameters:
\begin{equation}
\partial ^{i_{3}}\zeta _{i_{3},\nu _{1}\nu _{2}\nu _{3}} =0\text{ .}
\label{gauge-fix zeta}
\end{equation}
After imposing these gauge conditions all parameters $\xi$ can be solved for without any ambiguity, i.e.~there is
no gauge symmetry acting on the parameters left. This
leaves us with 490--315 = 175 degrees of freedom represented by the
$a,b,c,d,e$ components defined in eq.~\eqref{canonical decomp.}:
\begin{equation}\label{counting}
a:\ 50\,,\hskip .7truecm
b:\ 35\,,\hskip .7truecm
c:\ 35\,,\hskip .7truecm
d:\ 5\,,\hskip  .7truecm
e:\ 50\,.
\end{equation}

Using the canonical decomposition \eqref{canonical decomp.} we next calculate the different components of the Einstein tensor
\eqref{diffinv} and the Cotton tensor \eqref{Ctensor}. Substituting these results into the action \eqref{7Daction}, one obtains, after a lengthy calculation which we shall not repeat here, the following expression for the action \eqref{7Daction}:
\begin{eqnarray}
I &=& \int d^7x\ \bigg\{
-\frac{1}{2}\left( 3!\right) ^{4}b^{i_{2}i_{3},j_{2}j_{3}}\left( \nabla
^{2}\right) ^{2}\left( {\hat a}_{i_{2}i_{3},j_{2}j_{3}}+4\Box {\hat e
}_{i_{2}i_{3},j_{2}j_{3}}\right)   \notag \\ [.2truecm]
&&-\left( 3!\right) ^{4}m\,\hat {a}^{i_{2}i_{3},j_{2}j_{3}}\left( \nabla
^{2}\right) ^{2}\hat {e}_{i_{2}i_{3},j_{2}j_{3}}-\left( 3!\right)
^{4}m\,b^{i_{2}i_{3},j_{2}j_{3}}\left( \nabla ^{2}\right)
^{2}b_{i_{2}i_{3},j_{2}j_{3}}  \notag \\ [.2truecm]
&&-\frac{3}{4}\left( 3!\right) ^{4}m\,\bar{a}^{i_{3},j_{3}}\left( \nabla
^{2}\right) ^{2}\bar{e}_{i_{3},j_{3}}-10\left( 3!\right) ^{4}m\,a\left(
\nabla ^{2}\right) ^{2}e  \notag \\ [.2truecm]
&&-\frac{3}{10}\left( 5!\right) m\,c^{j_{1}j_{2},i_{2}}\nabla
^{2}c_{j_{1}j_{2},i_{2}}+\frac{9}{2}\left( 2!4!\right) m\,d^{j_{1}}\nabla
^{2}d_{j_{1}} \bigg\}\text{ .}
\end{eqnarray}
Here we have used the following decomposition of $a$  in terms of a traceless part $\hat {a}$, single traces $\bar a$ and double traces $a$:
\begin{eqnarray}
a_{i_{2}i_{3},j_{2}j_{3}} &=&{\hat a}_{i_{2}i_{3},j_{2}j_{3}} \ +  
\left\{ \left( \eta _{i_{2}j_{2}}-\frac{\partial _{i_{2}}\partial _{j_{2}}}{
\nabla ^{2}}\right) \bar{a}_{i_{3},j_{3}}\right.  \ + \notag \\
&&\ \ \ \ \ \ \left. +\left( \eta _{i_{2}j_{2}}\eta _{i_{3}j_{3}}-\frac{
\partial _{i_{2}}\partial _{j_{2}}}{\nabla ^{2}}\eta _{i_{3}j_{3}}-\eta
_{i_{2}j_{2}}\frac{\partial _{i_{3}}\partial _{j_{3}}}{\nabla ^{2}}\right)
a\right\}_{\text{a.s.}}
\end{eqnarray}
and we used a similar decomposition  for $e$.

Finally, after making the field redefinitions
\begin{eqnarray}\label{redefinitions}
\hat {a}_{i_{2}i_{3},j_{2}j_{3}} &=&\tilde{a}_{i_{2}i_{3},j_{2}j_{3}}-
\frac{2}{m}\Box b_{i_{2}i_{3},j_{2}j_{3}}\text{ ,} \hskip .7truecm
\hat {e}_{i_{2}i_{3},j_{2}j_{3}} =\tilde{e}_{i_{2}i_{3},j_{2}j_{3}}-
\frac{1}{2m}b_{i_{2}i_{3},j_{2}j_{3}}\text{ ,}
\end{eqnarray}
we obtain the following expression for the action:
\begin{eqnarray}
I &=&\int d^7x\ \bigg\{\frac{1}{m}\left( 3!\right)
^{4}b^{i_{2}i_{3},j_{2}j_{3}}\left( \nabla ^{2}\right) ^{2}\left( \Box
-m^{2}\right) b_{i_{2}i_{3},j_{2}j_{3}}  \notag \\ [.2truecm]
&&-\left( 3!\right) ^{4}m\,\tilde{a}^{i_{2}i_{3},j_{2}j_{3}}\left( \nabla
^{2}\right) ^{2}\tilde{e}_{i_{2}i_{3},j_{2}j_{3}}  \notag \\ [.2truecm]
&&-\frac{3}{4}\left( 3!\right) ^{4}m\,\bar{a}^{i_{3},j_{3}}\left( \nabla
^{2}\right) ^{2}\bar{e}_{i_{3},j_{3}}-10\left( 3!\right) ^{4}m\,a\left(
\nabla ^{2}\right) ^{2}e  \notag \\ [.2truecm]
&&-\frac{3}{10}\left( 5!\right) m\,c^{j_{1}j_{2},i_{2}}\nabla
^{2}c_{j_{1}j_{2},i_{2}}+\frac{9}{2}\left( 2!4!\right) m\,d^{j_{1}}\nabla
^{2}d_{j_{1}}\bigg\}\text{ .}
\end{eqnarray}
This form of the action shows that only the $b$ components  propagate and, according to eq.~\eqref{counting}, they do describe, unitarily, 35 degrees of freedom which transform as the ${\bf 35}^+$ of the SO(6) little group. Note that these degrees of freedom are not only described by the $b$-components
of $h$ but also, due to the redefinitions \eqref{redefinitions}, by the ${\hat a}$- and ${\hat e}$-components.
Replacing $m$ by $-m$ in the above action, we see that, after changing
the overall sign of the action, we again obtain 35 degrees of freedom. These degrees of freedom transform as the ${\bf 35}^-$ of the SO(6) little group. They are described by a different set of components of $h$ than the ${\bf 35}^+$ degrees of freedom due to the fact that one should also replace $m$ by $-m$ in the redefinitions \eqref{redefinitions}.

\section{Discussion}

We showed how the 3D TMG model, at the linearized level, can be extended beyond three dimensions to a free parity-odd Topologically Massive Gauge theory for a
``spin-2'' particle. We worked out the case of a massive ``spin-2'' particle in 7D; similar models exist in $4k-1$ dimensions for $k=3,4,5, \cdots$.
The construction of the model is based on the factorization of the Klein-Gordon operator in $4k-1$ dimensions, when acting on forms
of rank $2k-1$, in terms of two first-order operators.

A similar generalization of the parity-even 3D NMG model exists but in that case there are more extensions possible. For instance, a 4D extension
exists without a corresponding parity-breaking topological version \cite{Bergshoeff:2012ud}. In 7D there are three different extensions: one is
based on the same Young tableau \eqref{youngsymm} that we used for the  topological model constructed in this letter
and
one  is based on the dual of the spin connection, like in the 4D extension of \cite{Bergshoeff:2012ud}.\,\footnote{In 7D this corresponds to a description
in terms of a two-column Young tableau with height 5 and 1, respectively.} The third model is based on a description
in terms of a 2-column Young tableau of height 4 and 2, respectively. All these extensions have in common that the number of boxes $\#_{\text{boxes}}$ in the
two-column Young tableaux described by $h$ is given by
\begin{equation}
\#_{\text{boxes}} = D-1\,.
\end{equation}
One can show that this property guarantees that the index
structure of the double dual of the curvature tensor $R(h)$, which we have called the ``Einstein tensor'' $G(h)$,
is the same as that of $h$. This is a crucial property that enables one to
integrate the higher-derivative equations of motion to an action.

It is not difficult to write down the parity-even massive ``spin-2'' model
based on the Young tableau \eqref{youngsymm}. Starting from the corresponding generalized FP equations one ends up, after boosting up the derivatives, with the following action:
\begin{equation}
I [h] =\int d^{7}x\left\{ \frac{1}{72}h^{\bar{\mu},\bar{\nu}
}\varepsilon _{\bar{\mu}}{}^{\alpha \bar{\rho}}\partial _{\alpha }C_{\bar{
\rho},\bar{\nu}}\left( h\right) -\frac{1}{2}m^{2}h^{\bar{\mu},\bar{\nu}}G_{
\bar{\mu},\bar{\nu}}\left( h\right) \right\} \text{ ,}  \label{7Dactione}
\end{equation}
where $C_{\bar\mu,\bar\nu}(h)$ is the Cotton tensor, see eq.~\eqref{Ctensor}, and $G_{\bar\mu,\bar\nu}(h)$ is
the Einstein tensor, see eq.~\eqref{diffinv}. This action is the parity-even version of the action \eqref{7Daction}.
A canonical analysis, like the one we performed in section 3, shows that this model describes 70 ``spin-2'' states.

It is interesting to consider the massless limit of the models \eqref{7Daction} and \eqref{7Dactione}. A canonical analysis
shows that in the case of the parity-odd topological model \eqref{7Daction} the massless limit describes zero degrees of freedom while
for the parity-even model  \eqref{7Dactione} one ends up with 35 massless ``spin-2'' states which transform as the
${\bf 35}$ of the massless little group SO(5). The result
for the parity-odd model is similar to what happens for the 3D TMG model while the result for the parity-even model
resembles the parity-even cases in 3D  \cite{Deser:2009hb}
 and 4D \cite{Bergshoeff:2012ud}.

The crucial question remains whether the extensions we discussed in this letter are curiosities of the linearized approximation or whether
one can go beyond the linearized approximation and introduce non-trivial interactions. This is a non-trivial issue in view of the fact that we
are using non-standard representations to describe the massive ``spin-2'' particle. Perhaps, a slightly easier question to ask
is whether one can introduce interactions for only the mass term, i.e.~the term with two derivatives. For both the
parity-odd model \eqref{7Daction} and the parity-even model \eqref{7Dactione} this term is given by
\begin{equation}\label{trivial7D}
I\left[h\right] =\int d^{7}x\left\{
\frac{1}{2}h^{\bar{\mu},\bar{\nu}}G_{
\bar{\mu},\bar{\nu}}\left( h\right) \right\} \text{ .}
\end{equation}
This term by itself leads to the equation of motion $G(h)=0$ and therefore does not describe any degree of freedom, as
one would expect from a mass term. Given that there are no propagating degrees of freedom one might hope that it
will be an easier task to construct interactions.

The model \eqref{trivial7D} is the 7D version of the 3D gravity
action that neither describes any degree of freedom. The 3D gravity
action has the interesting feature that it can be reformulated as a
Chern-Simons (CS) action \cite{Achucarro:1987vz,Witten:1988hc}. In
order to achieve this, one must use a first-order formalism with the
Dreibein $e_\mu{}^a$ and spin-connection $\omega_\mu{}^a$ as
independent fields. Writing $e_\mu{}^a = \delta_\mu{}^a+ h_\mu{}^a$
this 3D CS action is at the linearized level given by
\begin{equation}\label{CSaction}
I_{\text{CS}}\left[h,\omega\right] =\int d^{3}x\, \varepsilon^{\mu\nu\rho}\left\{\omega_\mu{}^a\partial_\nu h_\rho{}^b\eta_{ab} - \frac{1}{2}\omega_\mu{}^a \delta_\nu{}^b \omega_\rho{}^c\varepsilon_{abc}\right\}\,.
\end{equation}
It is invariant under the linearized Lorentz transformation
\begin{equation}\label{3DL}
\delta h_{\mu a} = \Lambda_{\mu a}\,,\hskip 2 truecm \delta
\omega_\mu{}^a = -\frac{1}{2}\varepsilon^{abc}\partial_\mu
\Lambda_{bc}\text{ ,}
\end{equation}
for anti-symmetric parameters $\Lambda_{\mu a} = -\Lambda_{a\mu}$.
These linearized gauge transformations can be  fixed by imposing the gauge-fixing condition $h_{\mu a} = h_{a \mu}$.
One then obtains a first-order action in terms of $\omega_\mu{}^a$ and a symmetric tensor $h_{\mu\nu}$.
One of the reasons that this action can be extended to include interactions is that the Kronecker delta $\delta_\alpha{}^b$,
 occurring in the action \eqref{CSaction}, is in the same representation as the Dreibein $e_\mu{}^a$ and, therefore, can become part of this
  Dreibein at the non-linear level. The interactions are then determined by introducing the non-Abelian CS structure, dictated by the Lorentz structure of the different gauge fields.

It turns out that a similar first-order formulation exists of the model defined by the action \eqref{trivial7D}
in terms of  two fields $h_{\bar\mu,\bar\nu}$ and  $\omega_{\bar\mu,\bar\nu}$ which both have the symmetry properties corresponding
to the Young tableau
\begin{equation}\label{istr}
\Yvcentermath1
{\tiny \yng(1,1,1)}\otimes {\tiny
\yng(1,1,1)}\,.
\end{equation}
Similar to \cite{Skvortsov:2008sh}, at the quadratic level such a
first-order action  can be written in the following  form
\begin{equation}
I [ h,\omega ] =\int d^{7}x\ \varepsilon ^{\bar{\mu}\alpha
\bar{\nu }}\left\{ \omega _{\bar{\mu},}{}^{\bar{\rho}}\partial
_{\alpha }h_{\bar{\nu}, \bar{\rho}}-\frac{1}{72}\omega
_{\bar{\mu},}{}^{\bar{\sigma}}\delta _{\alpha }{}^{\beta }\omega
_{\bar{\nu},}{}^{\bar{\tau}}\varepsilon _{\bar{\sigma} \beta
\bar{\tau}}\right\} \text{ .}  \label{action h omega 2}
\end{equation}
This action  has a gauge invariance under a ``generalised'' linearized Lorentz transformation,
with parameters $\Lambda_{\mu_1\mu_2,\nu_1\nu_2\nu_3\nu_4}$, given by
\begin{eqnarray}\label{Lorentz-like transf. h}
\delta h_{\bar{\mu},\bar{\nu}}&=&\Lambda _{\lbrack \mu _{1}\mu
_{2},\mu _{3}]\nu _{1}\nu _{2}\nu _{3}}\,, \nonumber\\ [.2truecm]
\delta \omega _{\bar{\rho},}{}^{\bar{\mu}} &=&\varepsilon
^{\bar{\mu}\alpha \bar{\nu}}\partial _{\alpha }\Lambda _{\nu _{1}\nu
_{2},\nu _{3}\rho _{1}\rho _{2}\rho _{3}}-\frac{1}{4}\delta
_{\bar{\rho}}^{\bar{\mu} }\varepsilon ^{\bar{\sigma}\alpha
\bar{\nu}}\partial _{\alpha }\Lambda _{\nu _{1}\nu _{2},\nu
_{3}\sigma _{1}\sigma _{2}\sigma _{3}}   \\ [.2truecm] &&+\left\{
-\frac{9}{2}\delta _{\rho _{1}}^{\mu _{1}}\varepsilon ^{\sigma
_{1}\mu _{2}\mu _{3}\alpha \bar{\nu}}\partial _{\alpha }\Lambda
_{\nu _{1}\nu _{2},\nu _{3}\sigma _{1}\rho _{2}\rho _{3}}+3\delta
_{\rho _{1}\rho _{2}}^{\mu _{1}\mu _{2}}\varepsilon ^{\sigma
_{1}\sigma _{2}\mu _{3}\alpha \bar{\nu}}\partial _{\alpha }\Lambda
_{\nu _{1}\nu _{2},\nu _{3}\sigma _{1}\sigma _{2}\rho _{3}}\right\}
_{\text{a.s.}}\text{ .}\nonumber
\end{eqnarray}
In effect, the $\Lambda$-transformation represents three independent
gauge transformations whose parameters are given by the following
Young tableaux:
\begin{equation}\label{gparameters}
\Yvcentermath1
{\tiny \yng(1,1)}\,\otimes {\tiny
\yng(1,1,1,1)}\ = \ {\tiny \yng(1,1,1,1,1,1)}\ \oplus\ {\tiny \yng(2,1,1,1,1)}\ \oplus \ {\tiny \yng(2,2,1,1)}\ \,.
\end{equation}
The gauge transformations \eqref{Lorentz-like transf. h} are the generalization of the 3D Lorentz transformations \eqref{3DL}.

It is easy to see that the action \eqref{action h omega 2} is equivalent to \eqref{trivial7D}. One first imposes the condition
\begin{equation}
h_{\bar\mu,\bar\nu} = Y_{[3,3]}\,  h_{\bar\mu,\bar\nu}
\end{equation}
to fix the gauge transformations \eqref{Lorentz-like transf. h}.
Next, one uses the equation of motion for $\omega_{\bar\mu,\bar\nu}$
to solve for $\omega_{\bar\mu,\bar\nu}$ in terms of $h_{\bar\mu,\bar\nu}$:
\begin{equation}\label{solo}
\omega_{\bar\mu,\bar\nu} = \epsilon_{\bar\nu}{}^{\alpha\bar\rho}\partial_\alpha h_{\bar\rho,\bar\mu}\,.
\end{equation}
Note that this equation implies that $\omega_{\bar\mu,\bar\nu}$ is traceless, i.e.~$\eta^{\mu\nu}\omega_{\bar\mu,\bar\nu}=0$.
Substituting this solution back into \eqref{action h omega 2} the two terms in \eqref{action h omega 2}
coincide and become identical to the single term in \eqref{trivial7D} with the Einstein tensor given in eq.~\eqref{diffinv}.

The gauge-invariant first-order formulation we have obtained at this point resembles the 3D CS structure. There are, however, also  important differences. First of all, it is not clear how to introduce  in the 7D case the notion of  flat and curved indices, thereby anticipating a possible CS-like structure. A related issue is that we are working now with tensors instead of gauge vectors. It is
not obvious how to introduce non-Abelian structures for these tensors. The structure we have obtained so far suggests an extension of CS terms for vectors to a ``generalised CS'' structure for a non-Abelian version of free differential algebras.
An alternative approach to introduce interactions could be to use a bi-metric formulation. One metric describes the massive spin-2 particle and is used  to absorb the $h_{\bar\mu,\bar\nu}$ field, while the other metric is a reference metric that can be used to absorb the Kronecker delta that occurs in the second term of \eqref{action h omega 2}.
For now, we leave these possibilities as intriguing open issues.

\section*{Acknowledgements} We thank Paul Townsend for useful discussions and for pointing out reference \cite{Lu:2010sj} to us. YY wishes to thank Andrea Borghese, Giuseppe Dibitetto, Jose Juan Fernandez-Melgarejo, Teake Nutma and Diederik Roest for discussions on group theory and useful software.
The work of JR is supported by the Stichting Fundamenteel Onderzoek der Materie
(FOM). The work of MK and YY is supported by the Ubbo Emmius Programme administered
by the Graduate School of Science, University of Groningen.
We acknowledge the frequent use of the  software Cadabra \cite{Peeters:2006kp} to perform Young projection calculations.


\providecommand{\href}[2]{#2}\begingroup\raggedright\endgroup

\end{document}